\documentclass[pra,aps,groupedaddress,twocolumn,floatfix,nofootinbib]{revtex4}

\usepackage{graphicx}
\newcommand{\beq}{\begin{equation}}
\newcommand{\eeq}{\end{equation}}
\newcommand{\beqa}{\begin{eqnarray}}
\newcommand{\eeqa}{\end{eqnarray}}

\providecommand{\expect}[1]{\langle#1\rangle}

\newcommand{\ket}[1]{\mbox{$ | #1 \rangle $}}

\def\opone{\leavevmode\hbox{\small1\normalsize\kern-.33em1}}

\begin{document}

\title{
Quantum correlations in Newtonian space and time:\\
arbitrarily fast communication or nonlocality}
\author{Nicolas Gisin}

\date{\small \today}

\begin{abstract}
We investigate possible explanations of quantum correlations that satisfy the principle of continuity, which states that everything propagates gradually and continuously through space and time. In particular, following [J.D. Bancal et al, Nature Physics 2012], we show that any combination of local common causes and direct causes satisfying this principle, i.e. propagating at any finite speed, leads to signalling. This is true even if the common and direct causes are allowed to propagate at a supraluminal-but-finite speed defined in a Newtonian-like privileged universal reference frame. Consequently, either there is supraluminal communication or the conclusion that Nature is nonlocal (i.e. discontinuous) is unavoidable.\\

It is an honor to dedicate this article to Yakir Aharonov, the master of quantum paradoxes.
\end{abstract}

\maketitle

\section{Introduction}\label{intro}
Correlations cry out for explanations \cite{bellbook}. This is true in all sciences, from correlations between measurement results in quantum physics to correlations between earthquakes and tsunamis in geophysics, and correlations between tides and the moon's positions in classical physics, to name but a few examples. Once a correlation has been identified, the next task of science consists in developing a theoretical model explaining the correlation. Such models take the form of a story supported by mathematical equations. Particularly challenging is the search of an explanation for quantum correlations when considering several measurements per party on two or more distant systems initially in an entangled state.

In all sciences besides quantum physics, all correlations are explained by a combination of only two basic mechanisms. Either a first system influences a second one, i.e. Direct Causation (DC), as for example the earthquake that causes the tsunami. Or the correlated events share a local Common Cause (CC) in their common past as two readers of this text whose readings are highly correlated. Sometimes the common or direct causes may be subtle and not easy to detect, as twins that look extraordinarily alike thanks to common genes (local variables, i.e. CC), or as
one's yawning triggers others to yawn, thanks to delicate influences (i.e. DC).

Many correlations involve a combination of the two basic mechanisms, common and direct causes, like for instance the correlations between hockey players: they trained together, hence share common causes, and, during games, influence each other.

Formally a correlation between two parties A and B is a conditional probability distribution $p(a,b|x,y)$, where $a,b$ denote the measurement results collected by A and B, and $x,y$ the measurement settings freely (i.e. independently from each other and from all CC and DC) chosen by A and B, respectively. This generalizes straightforwardly to n parties. If A's marginal $p(a|x,y)\equiv\sum_b p(a,b|x,y)$ depends explicitly on B's choice $y$, then A can get information about B's choice by merely observing her local statistics. This is called {\it signalling}. The no-signalling principle states that A's marginal is independent of B's choice, $p(a|x,y)\equiv \sum_b p(a,b|x,y)=p(a|x)$, and B's marginal is independent of A's choice, $p(b|x,y)=p(b|y)$. Note that all physical communication should be carried by some physical object (atoms, photons, energy, waves, etc). Hence, assuming only local Common Causes carried by the (localized) physical systems in Alice and Bob's hands, signalling would be non-physical communication. But Direct Cause may allow signalling as discussed in section(\ref{nosignalling}).

This paper is organized as follows. In the next section, we present the intuition behind our result \cite{Pironioetal}. Next, in section \ref{vcausality}, we define formally $v$-causal models. Then, before presenting the main result in section \ref{NoInfluenceNorLocalVariables}, we analyze the case of DC (without additional variables) in section \ref{NoInfluence}. Finally, we discuss experiments that could test our results in section \ref{experiment} and discuss the interpretation of our results.

\section{Explanations of correlations}\label{explanations}
First attempts at explaining correlations between distant quantum measurement results assumed that the source producing the entangled quantum systems produces additional variables, hidden to today's physics, which would locally (i.e. continuously) determine the probabilities of the measurement results. This would provide a local Common Cause explanation.
Such local hidden variable models must obey the famous Bell inequalities. But quantum theory predicts and experiments confirm that Bell inequalities can be violated; hence all explanations based only on local common causes have been experimentally refuted\footnote{up to some combinations of loopholes that seem highly implausible; however, this being science, this logical possibility should be addressed experimentally.}.

Direct Cause explanations of quantum correlations received relatively little attention, compared to CC explanations (up to some noticeable exceptions, in particular Eberhard who proposed an explicit model already in 1989 \cite{Eberhard}). This is due to the fact that Bell inequality violations have been convincingly demonstrated between space-like separated measurements \cite{locality1,locality2,locality3}, hence a DC explanations would require influences that propagate faster than light.

The assumption of faster than light influences does not respect the spirit of relativity. However, the assumption of a universal privileged reference frame with respect to which a faster than light influence can be defined, is not in contradiction with relativity\footnote{One could also consider the history-fiction case that quantum theory would have been developed before the discovery of relativity. In such a case, quantum nonlocality would have been equally surprising and fascinating and physicists would naturally have been led to search for explanations of these extraordinary correlations in terms of delicate influences yet to be discovered.}. Think for example of the reference frame in which the micro-wave back ground radiation, residue of the big bang, is isotropic; our Earth propagates with respect to this universal frame at the well defined speed of 369 km/s in a direction known at each moment \cite{microRefFrame}.

There is thus no definite reason not to investigate the possibility of explaining quantum correlation with a combination of DC and CC. Actually, many authors who thought seriously about quantum non-locality noticed that correlation between distant events strongly suggest that ``something is going on behind the scene", using John Bell's words \cite{BellNouvelleCuisine,BellHiddenComm}. David Bohm and Basil Hiley, for example, have been very explicit when writing ``it is quite possible that quantum nonlocal connections might be propagated, not at infinite speeds, but at speeds very much greater than that of light. In this case, we could expect observable deviations from the predictions of current quantum theory (e.g. by means of a kind of extension of the Aspect-type experiment)" \cite{BohmHiley}. Let us also note that most (non relativistic) text books tell a story like ``first measurement collapses the entire wavefunction, hence changes (influences) the state of all systems entangled with the measured system". Consequently, it is good scientific practise to study the assumption that quantum correlations are caused by faster than light influences propagating in a hypothetical universal privileged reference frame and analyze its consequences.

We call {\bf $v$-causal} all explanations that combine local Common Causes and Direct Causes where the influence (describing the direct cause) propagates at a supraluminal-but-finite speed $v$ defined in a hypothetical universal privileged reference frame: $c< v<\infty$. Note that such a universal privileged reference frame would be quite similar to Newton's space and time, but with a given fixed maximal velocity $v$. It is thus quite familiar to physicists\footnote{Though this strongly contrasts with ideas in quantum gravity where space-time is sometimes thought of as an emergent concept, as e.g. in loop quantum gravity.}, see Figure 1. When two events can be connected by a hidden influence at speed $v$, we say that they are {\bf $v$-connected}; otherwise we say that the events are {\bf not-$v$-connected}.

\begin{figure}
\centering
\includegraphics[width=9cm]{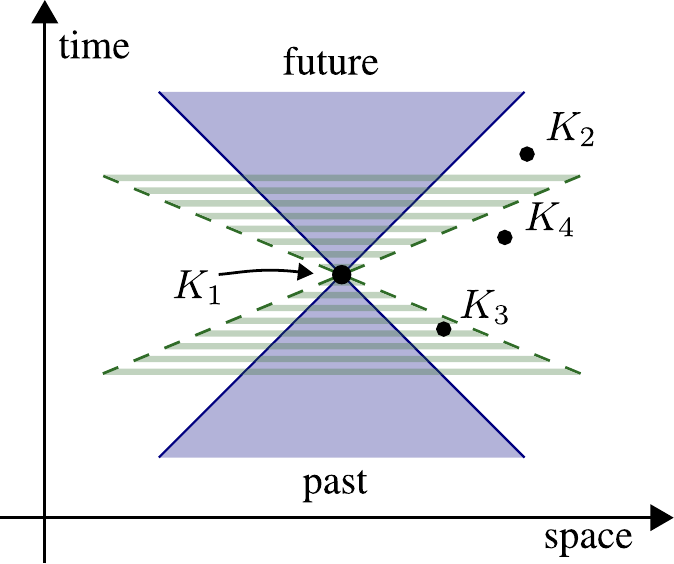}
\caption{Space-time diagram in the privileged reference frame.  The shaded light cone is delimited by solid lines. Points inside the $v$-cone (hatched), e.g. K2 and K3, are $v$-connected to K1; while points outside the $v$-cone, like K4, are not-$v$-connected to K1. (Taken with permission from Nature \cite{Pironioetal}).}\label{fig1}
\end{figure}

The kind of experiment that Bohm and Hiley had in mind to test such a DC or $v$-causal explanation is quite intuitive: if the influence carrying the DC propagates at finite speed, it should be possible to arrange an experiment between distant quantum systems with good enough synchronization (in the universal privileged reference frame), so that the influence doesn't arrive on time to establish the correlation. Thus, in such situations, the measured correlation should necessarily be local, i.e. satisfy all Bell inequalities, even in cases where quantum theory predicts a violation of some Bell inequality. Hence, $v$-causal explanations can't reproduce all quantum predictions. Accordingly, they can be tested experimentally against quantum theory.

Such experiments face two intrinsic difficulties. First, since today we don't know the hypothetical privileged reference frame, it is not clear in which reference frame the synchronization should be optimized. Indeed, if two events are simultaneous in one frame, e.g. the laboratory frame, then, according to special relativity, they are not simultaneous with respect to others frames, e.g., to the cosmic microwave background radiation frame. Second, within an assumed privileged reference frame, perfect synchronization is impossible in practice; hence if nonlocal correlations are observed, this only sets a lower bound on the speed of the hypothetical hidden influence. Nevertheless, experiments have been carried out, setting stringent lower bounds of this speed, assuming the lab frame \cite{MovingObs}, the microwave background radiation frame \cite{pla00} and even scanning all possible privileged reference frames \cite{Salart,Cocciaro10}. These experiments have excluded speeds up to about 50'000 times the speed of light.

At this point the case for a definite experimental test of DC explanation may seem quite hopeless: two-party experiments can only  hope to increase the lower bound of the speed of the hypothetical hidden influence or to find the breakdown of quantum theory. But in 2002 Valerio Scarani and myself noticed that the situation changes dramatically when analyzing situations with more than two parties \cite{pla02,brazilian}. The original scenario we considered involves 3 parties (see also Ryff \cite{RyffTriangle} whose argument is recalled in section \ref{NoInfluence}). The general idea is the following. If two out of all parties measure simultaneously, e.g Bob and Charlie are not-$v$-connected, then their correlation must be local. If moreover, the correlations between the other pairs of parties, those whose measurements are $v$-connected, allow one to guarantee that Bob and Charlie share nonlocal correlations, then one could infer a contradiction with any $v$-causal explanation without the need for any demanding synchronization. That one can infer the nonlocality between Bob and Charlie without ever measuring them in the same run of an experiment is quite counterintuitive, though it is known that sometimes one can infer a property of some quantum state or probability distribution from only the knowledge of some of their marginals \cite{Linden,Wurflinger12}.

The next step was made by Stefan Wolf and colleagues who introduced the concept of transitivity of nonlocality \cite{zurich}. They showed that, assuming only no-signalling, there are examples of 3-party correlations, $p(a,b,c|x,y,z)$, such that if both marginals A-B and A-C are nonlocal, then the third marginal B-C is necessarily also nonlocal. This beautifully illustrates the idea Scarani and myself had in 2002. But unfortunately, Wolf and colleagues's example uses correlations that can't be achieved with measurements on quantum systems and, today, no quantum example of transitivity of nonlocality has been found. This is why the example we present in this paper doesn't use the concept of transitivity of nonlocality, but the theorem \cite{Pironioetal} recalled in section \ref{NoInfluenceNorLocalVariables}.

To conclude this introduction let us consider some consequences of the assumption that $v$-causality is the explanation of all quantum correlations. As already mentioned, this would imply that some predictions of quantum theory are wrong: if two events are not-$v$-connected, then their correlation would be local even in cases where quantum theory predicts a violation of some Bell inequality. But could this departure from quantum predictions be used to communicate, in particular to communicate faster than light? In the 2-party case, Alice and Bob could arrange to be just at the border of being $v$-connected. So, if Bob makes his measurement early enough, the hidden influence doesn't arrive on time and they observe local correlations; but if Bob delays a little bit his measurement, then the influence arrives on time and they observe quantum correlations. This, however, can't be used by Alice and Bob to communicate. Indeed, their local statistics would be identical in both cases, whether the hidden influence arrives on time or not; it is only later, once they compare their data, that Alice and Bob can notice whether or not they violated some Bell inequality. Consequently, with only two parties, the hidden influence could remain hidden for ever: there would be a hidden layer at which faster than light hidden influences carry Direct Causes and thus establish correlations that appear nonlocal, but at our higher level nothing travels faster than light. In this paper, following \cite{Pironioetal}, we prove that such a peaceful coexistence between relativity and faster than light hidden influences can't exist. But for this we'll need to consider more than two parties.

\section{$v$-causality}\label{vcausality}
In this section we define formally local Common Cause, Direct Cause and $v$-causal explanations. Readers who feel they understand CC, DC and $v$-causality may like to jump to section \ref{NoInfluence}.

Consider a 2-party scenario, denoted Alice and Bob, with measurement settings $x$ and $y$ and measurement results $a$ and $b$, respectively. The generalization to more parties is straightforward, as summarized at the end of this section. The conditional probability distribution, or in short correlation, $p(a,b|x,y)$, is the probability of results $a,b$ when the settings $x,y$ are chosen.

A pure {\bf local Common Cause} explanation of $p(a,b|x,y)$ assumes additional variables, traditionally labeled $\lambda$, such that:
\beq\label{BellLocality}
p(a,b|x,y)=\sum_\lambda~\rho(\lambda)p(a|x,\lambda)p(b|y,\lambda)
\eeq
where $\rho(\lambda)$ denotes the probability that the additional variable assumes the value $\lambda$ (note that $\lambda$ may include the quantum state $\rho$). For a justification see, e.g. \cite{bellbook,Mermin,maudlin,norsen,GisinFoundPhys}. In a $v$-causal model, the information carried by the variable $\lambda$ propagates gradually and continuously from some common $v$-past of Alice and Bob. If $v$ would be the speed of light, this would merely be the usual intersection between the past light cones. But here the common $v$-past is the intersection of wider, more open, cones, see Fig.1. Important in a common cause explanation is that $p(a|x,\lambda)$ doesn't depend on $y$ and symmetrically $p(b|y,\lambda)$ is independent of $x$. Hence all correlations are due to the common local variable $\lambda$.

A pure {\bf Direct Cause} explanation of $p(a,b|x,y)$ assumes that there is an absolute time ordering of the events at Alice and Bob (defined in the hypothetical universal privileged reference frame). For example, assume Alice is first to chose her measurement settings $x$ and collect her result $a$. Direct cause\footnote{Standard text book descriptions of measurements collapsing the quantum state is an explicit example of a hidden influences explanation; however, in such descriptions the influence propagates at infinite speed. Hence it is more a direct action at a distance than an influence propagating in space and time. Note that because of the infinite speed, all parties are $v$-connected. Such descriptions also require a universal privileged reference frame.} assumes that as soon as Alice performed her measurement, a signal - which we call a hidden influence - informs the rest of the universe, in particular Bob, of her measurement setting $x$ and result $a$. In this case there are two possibilities:
\begin{enumerate}
\item The information reaches Bob's system before it produces the result $b$, i.e. Alice and Bob are $v$-connected. In this case:
    \beq
    p(a,b|x,y,v\text{-connected})=p(a|x)p(b|y,~x,a)
    \eeq
    For example, quantum correlations between $v$-connected events can be described as due to DC: $p(a|x)=Tr(A_a^x\rho_A)$ where $\rho_A$ Alice's partial trace quantum state and $A_a^x$ the projector representing her measurement, and $p(b|y,~x,a)=Tr(B_b^y\rho_a^x)$ where $\rho_a^x=\frac{A_a^x\rho A_a^x}{Tr(A_a^x\rho)}$ is Bob's reduced state that depends on Alice's measurement setting $x$ and result $a$. Note that in this case direct cause exactly reproduces the quantum prediction: $p(a,b|x,y, v$-connected$)=Tr(A_a^x\otimes B_b^y\cdot\rho)$.
\item Bob's system has to produce the result $b$ before the information carried by the hidden influences arrives from Alice's system, i.e. Alice and Bob are not-$v$-connected:
    \beqa
    & &p(a,b|x,y, \text{ not-}v\text{-connected})\nonumber\\
    &=&p(a|x)p(b|y)
    \eeqa
    In the case that Bob's probability depends only on his local quantum state $\rho_B=Tr_A(\rho)$, one has:
    \beqa
    & &p(a,b|x,y, \text{ not-}v\text{-connected})\nonumber\\
    &=& Tr(A_a^x\cdot\rho_A)\cdot Tr(B_b^y\cdot\rho_B)
    \eeqa
    In general, for entangled states $\rho$, this prediction differs from the quantum prediction.
\end{enumerate}

A {\bf $v$-causal} explanation of $p(a,b|x,y)$ combines additional local variables and hidden influences\footnote{The De-Broglie-Bohm pilot wave model is an explicit example of a $v$-causal explanation; however, in this model the influence propagates at infinite speed. Hence it is more a direct action at a distance than an influence propagating in space and time. Note that because of the infinite speed, all parties are $v$-connected, hence Bohm's model recovers all quantum predictions. This model also requires a universal privileged reference frame.}, all propagating at a speed $v$ (or lower) defined in the universal privileged reference frame. This frame defines an absolute time ordering, as for direct cause explanations. Here again one has to distinguish two possibilities depending on whether Alice and Bob are $v$-connected or not:
\begin{enumerate}
\item The information reaches Bob's system before it produces the result $b$, i.e. Alice and Bob are $v$-connected. In this case:
    \beqa
    p(a,b|x,y, v\text{-connected})\nonumber\\
    =\sum_\lambda \rho(\lambda) p(a|x\lambda)p(b|y,\lambda,~x,a)
    \eeqa
     Since we look for an explanation of quantum correlations, one expects that, in the case of $v$-connected events, quantum correlations are reproduced.
\item Bob's system has to produce the result $b$ before the information carried by the hidden influences arrives from Alice's system, i.e. Alice and Bob are not-$v$-connected:
    \beqa
    p(a,b|x,y, \text{ not-}v\text{-connected})\nonumber\\
    =\sum_\lambda \rho(\lambda) p(a|x,\lambda)p(b|y,\lambda)
    \eeqa
     where $\lambda$ includes the quantum state $\rho$. Consequently, in any $v$-causal model, unconnected events must satisfy all Bell inequalities.
\end{enumerate}

The generalization to an arbitrary number of parties should be straightforward: when a system undergoes a measurement it takes into account all the information it received, whether additional local variables or hidden influences, and sends out information about itself in all directions by hidden influences propagating at speed $v$. Since we are looking for an explanation of quantum correlations, one expects that, whenever possible, $v$-connected events reproduce quantum correlations. However, unconnected events necessarily produce local correlations, hence correlations that may differ from the quantum predictions. This constraint is what limits the power of $v$-causal explanations and makes experimental tests possible.

Note that the speed of light $c$ doesn't appear in the definitions of local Common Cause and Direct Cause, nor $v$-causality \footnote{nor does c appear in the definition of "Bell locality" (\ref{BellLocality}). Nevertheless, physicists have always been interested in tests of Bell inequalities between space-like separated events, i.e. between events not-$c$-connected. This illustrates that $v$-causal models were always in the back of the mind of those physicists, though with $v=c$.}.

\section{No direct cause explanation}\label{NoInfluence}
In this section we study the assumption that correlations between quantum measurement results are due to DC carried by hidden influences propagating at a finite but supraluminal speed $v$. More precisely, we consider hidden influence plus the usual quantum state, but no additional local variables. This section is greatly inspired by \cite{RyffTriangle} (note that Eberhard published a related argument also involving 3 parties \cite{Eberhard}).

Consider a 3-party scenario, Alice, Bob and Charlie, where Alice is far away from both Bob and Charlie. Bob and Charlie are relatively close to each other, but distant enough (in the hypothetical universal reference frame) so that they can synchronize their measurements well enough to be not-$v$-connected, see Fig. 2. Alice, Bob and Charlie know the relative positions of each other and at what time Bob and Charlie perform their measurements. Assume they share a GHZ state $\Psi=\ket{0,0,0}+\ket{1,1,1}$ and all measure $\sigma_z$. Quantum theory predicts that all three collect the same result: $a=b=c$. We shall see that if this correlation is due to some supraluminal hidden influence (without additional variables), then Alice could communicate faster than light to Bob and Charlie (Bob and Charlie need to collaborate).

\begin{figure}
\centering
\includegraphics[width=9cm]{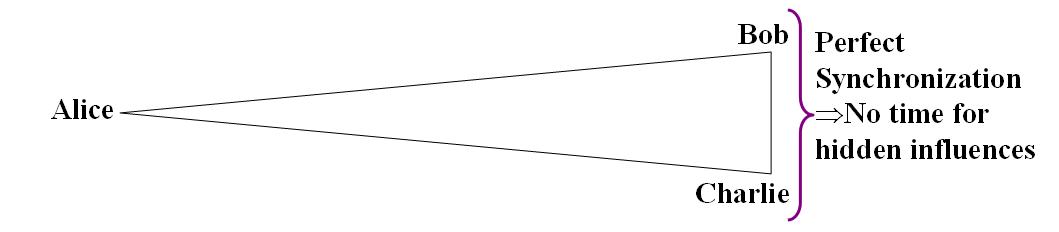}
\caption{Spacial configuration of the 3-party scenario discussed in section \ref{NoInfluence} to show that pure Direct Cause leads to signalling.}\label{fig2}
\end{figure}

The argument runs as follows. First, if Alice chooses to communicate ``yes", she performs her measurement early enough that the hidden influence arrives on time to Bob and Charlie. In this case the hidden influence tells Bob and Charlie's system which result $a$ Alice obtained, hence Bob and Charlie's system produce that same result: $b=a$ and $c=a$. Next, if Alice chooses to communicate ``no", she doesn't perform any measurement, or only too late for the hidden influence to arrive on time. In this case Bob and Charlie obtain random and independent results (recall that they are not-$v$-connected, hence their result are produced independently of each other), whence half the time $b\neq c$. Consequently, once Bob and Charlie compared their results (which they can do in a time very short relative to the time light would take to propagate from Alice to them), they can infer with good probability Alice's message.

This is faster than light communication from Alice to Bob-Charlie. By elongating the triangle the speed of this communication gets arbitrarily close to the speed $v$ of the hidden influence. Hence, the hidden influence doesn't remain hidden, but can be activated.

This simple example shows that with 3 parties one can activate the hidden influence, something impossible with only 2 parties. However, this example also shows that there is a simple way around the argument. Indeed, the correlation is a simple and local one: $a=b=c$. Hence, one could merely supplement the DC explanation with a shared random bit $r$ and assume that in the case the hidden influence doesn't arrive on time, all systems produce the result $r$. This motivates the investigation of $v$-causality, where DC is combined with additional local variables as explained in section \ref{vcausality} and analyzed in the next section.

\section{No $v$-causal explanation}\label{NoInfluenceNorLocalVariables}
At this stage of the search for an explanation of quantum correlations, local common causes and hidden influences are both individually excluded. The first one predicts Bell inequalities that have been violated, while the second one can't remain hidden as recalled in the previous section. Let us thus analyse the hypotheses that $v$-causality, i.e. an arbitrary combination of Direct and Common Causes, is the explanation of all correlations. This might sound bizarre. But quantum correlations are bizarre and there is simply no other type of explanations that satisfy the principle of continuity (we discuss this principle in more detail in section\ref{Newton}). This section is greatly inspired by \cite{Pironioetal}.

Consider the 4-party configuration of figure 3, represented in the hypothetical privileged reference frame. Alice, Bob, Charlie and Dave have a choice between two measurement settings, labeled $x,y,z$ and $w$ and collect binary results $a,b,c,d\in\{-1,+1\}$, respectively. Alice measures first, hence is not influenced by any of the other parties. Next, Dave measures at a time such that the hypothetical influence from Alice arrives on time to Dave.
Finally, Bob and Charlie measure quasi-simultaneously, i.e. Bob and Charlie are not-$v$-connected, but such that the hypothetical influences from Alice and Dave arrive on time both to Bob and to Charlie.

\begin{figure}
\centering
\includegraphics[width=9cm]{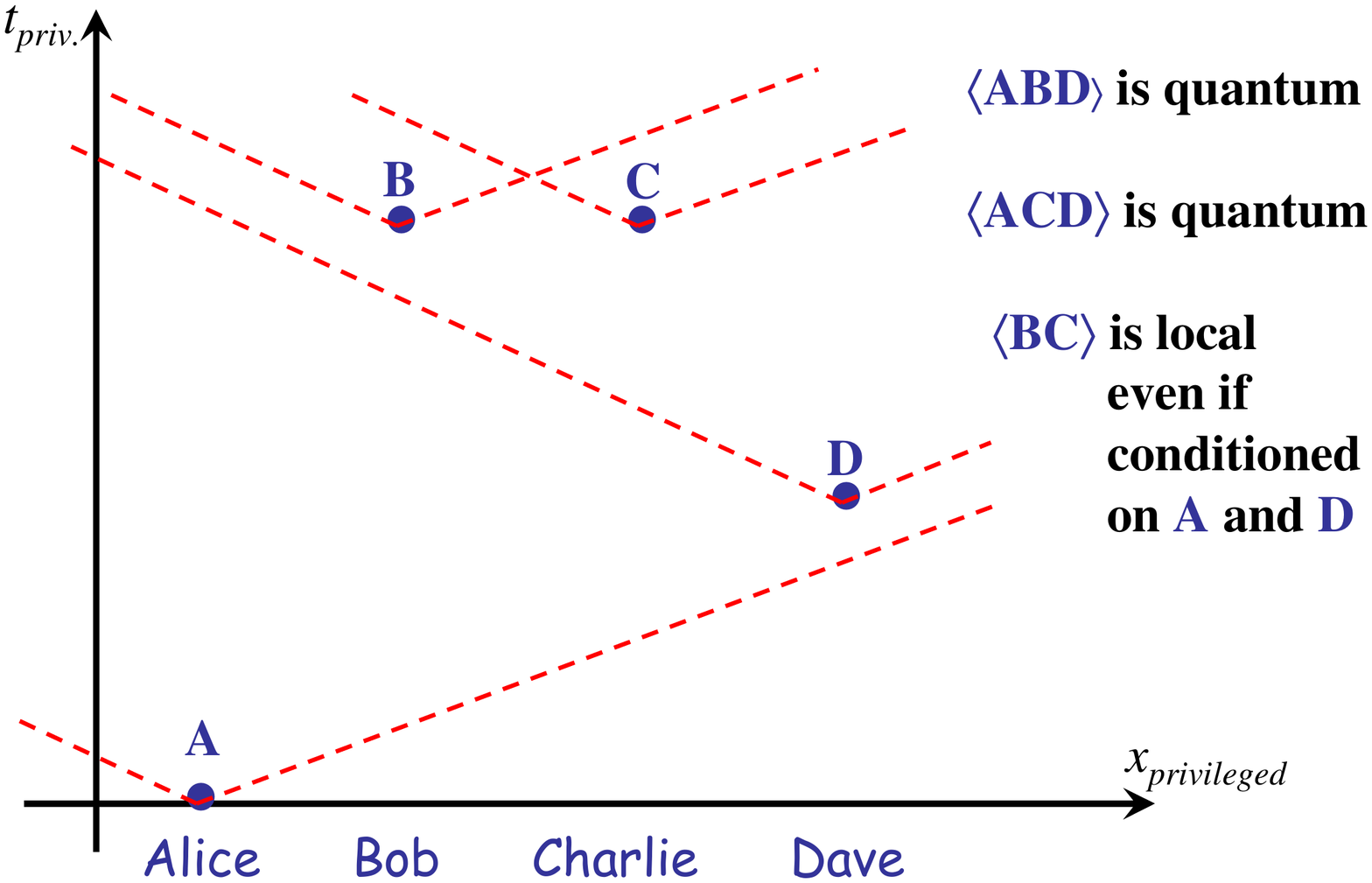}
\caption{Space-time configuration in the privileged reference frame of the 4-party scenario discussed in section \ref{NoInfluenceNorLocalVariables} to show that all $v$-causal models lead to signalling. (Taken with permission from Nature from \cite{Pironioetal}).}\label{fig3}
\end{figure}

If we were considering only DC, the joined probability would read:
\beqa\label{pDC}
&p&(a,b,c,d|x,y,z,w)= \\ \nonumber
&p&(a|x)\cdot p(d|w,x,a)\cdot p(b|y,x,a,w,d)\cdot p(c|z,x,a,w,d)
\eeqa
It is not difficult to see that the correlation (\ref{pDC}) leads to signalling from Alice to Bob-Charlie-Dave (who need to cooperate). But in this configuration, contrary to the triangular configuration of the previous section, we can exclude the possibility that additional variables allow one to avoid the activation of the hypothetical hidden influence.

The idea is to find an inequality satisfied by all no-signalling correlations where the not-$v$-connected parties are local with the following two properties:
\begin{enumerate}
\item all terms in the inequality involve only $v$-connected parties (hence, to evaluate the inequality one never has to measure in a same round of the experiment not-$v$-connected parties, one thus avoids the synchronization difficulty),
\item the n-party correlation can be violated by quantum correlations (i.e. quantum theory predicts a violation of the inequality).
\end{enumerate}

The technical difficulty of this strategy is that, first one has to study the intersection of the n-party no-signalling polytope with the local potytope of the not-$v$-connected parties. Next, one has to project this intersection polytope on the subspace of correlations containing only terms corresponding to $v$-connected parties.

This strategy can obviously not work with only two parties (both would either be $v$-connected or both not-$v$-connected). Hence, with my co-authors of \cite{Pironioetal} we spent a long time searching for an example involving 3 parties, one pair being not-$v$-connected and two pairs $v$-connected. But no example has been found, though the search continues, varying the number of inputs (measurements settings) and outcome for each party \cite{TomerMaster}. The breakthrough came when Jean-Daniel Bancal and Stefano Pironio had the courage to consider 4 parties in the configuration of figure 3. After heavy numerical search they found the following \cite{Pironioetal}.\\

{\bf Theorem}
Let $p(a,b,c,d|x,y,z,w)$ be a correlation, i.e. a conditional probability distribution, with binary inputs $x,y,z,w \in\{0,1\}$ and outcomes $a,b,c,d \in\{-1,+1\}$.\\
If
\begin{enumerate}
\item The correlation $p(a,b,c,d|x,y,z,w)$ is non-signalling, and
\item
$p(b,c|y,z,~a,x,d,w)$ is local\footnote{i.e. satisfy the Clauser-Horn inequality: $p(b=c=0|0,0,~a,x,d,w)+p(b=c=0|0,1,~a,x,d,w)+p(b=c=0|1,0,~a,x,d,w)-p(b=c=0|1,1,~a,x,d,w)-p(b=0|y=0,~a,x,d,w)-p(c=0|z=0,~a,x,d,w)\le 0$ and all its symmetric forms obtained by permuting the inputs and outcomes.} for all $a,x,d,w$,
\end{enumerate}
then $S\le7$, where
\beqa\label{ineqS}
S &=& - 3\expect{A_0} - \expect{B_0} - \expect{B_1} - \expect{C_0} - 3\expect{D_0} \nonumber\\
    & -& \expect{A_1B_0} - \expect{A_1B_1} + \expect{A_0C_0} \nonumber\\
    & +& 2\expect{A_1C_0} + \expect{A_0D_0} + \expect{B_0D_1} \nonumber\\
    & -& \expect{B_1D_1} - \expect{C_0D_0} - 2\expect{C_1D_1} \nonumber\\
    & +& \expect{A_0B_0D_0} + \expect{A_0B_0D_1} + \expect{A_0B_1D_0} \nonumber\\
    & -& \expect{A_0B_1D_1} - \expect{A_1B_0D_0} - \expect{A_1B_1D_0} \nonumber\\
    & +& \expect{A_0C_0D_0} + 2\expect{A_1C_0D_0} - 2\expect{A_0C_1D_1} \label{Sineq}
\eeqa
In (\ref{ineqS}) $\expect{A_1B_0}$ denotes the average of the product of Alice and Bob's outcomes when Alice chooses $x=1$ and Bob $y=0$ and similarly for the other terms.\\

The above inequality $S$ is remarkable because none of its 23 terms involves both Bob and Charlies, hence it can be evaluated without ever measuring Bob and Charlie in the same run of an experiment. Nevertheless,
\begin{enumerate}
\item Assuming no-signalling, a violation implies that Bob and Charlie share nonlocal correlations, i.e. correlations that can't be explained by Common Causes, and
\item Assuming that Bob and Charlie are local, as they are in any $v$-causal model, a violation implies that $p(a,b,c,d|x,y,z,w)$ is signalling.
\end{enumerate}

It is not difficult to check that the inequality $S\le 7$ can be violated by the following 4 qubit state \cite{Pironioetal}
\beqa\label{Psi}
\ket{\Psi}&=& \frac{17}{60}\ket{0000} + \frac{1}{3}\,\ket{0011} - \frac{1}{\sqrt{8}}\ket{0101} + \frac{1}{10}\,\ket{0110}\nonumber\\
          &+& \frac{1}{4}\ket{1000} - \frac{1}{2}\ket{1011} - \frac{1}{3}\ket{1101} + \frac{1}{2}\ket{1110}\label{Psi}
\eeqa
with the measurements
\beqa\label{meas}
\hat{A}_0= -U\sigma_xU^\dag &\hspace{1cm}& \hat{A}_1=U\sigma_z U^\dag \label{settings1}\\
\hat{B}_0=H &\hspace{1cm}& \hat{B}_1=-\sigma_x H \sigma_x \label{settings2}\\
\hat{C}_0=-\hat{D}_0=\sigma_z &\hspace{1cm}& \hat{C}_1=\hat{D}_1=-\sigma_x \label{settings3}
\eeqa
where $U=\cos(\frac{4\pi}{5})\sigma_z-\sin(\frac{4\pi}{5})\sigma_x$, the $\sigma$'s denote the Pauli matrices and
$H$ the Hadamard matrix. Quantum theory predicts for these state and measurement settings $S\approx7.2$.

Accordingly, the supraluminal hidden influence in any $v$-causal model can be activated. Indeed, in any $v$-causal model Bob and Charlie are local, hence, one can deduce from the quantum prediction that the 4-party correlation is signalling (recall that the 4-party correlation is not quantum, because Bob and Charlie are not-$v$-connected, only the 3-party marginals A-B-D and A-C-D are quantum, but this suffices to evaluate $S$).

Consequently, at least one of the four 3-party marginals depends on the fourth's input. Consider first the A-B-D 3-party marginal; since A-B-D are all $v$-connected, $p(a,b,d|x,y,w)$ is quantum and thus non-signalling (it doesn't depend on Charlie's input $z$). Moreover, the A-B-D correlation can't depend on Charlie's input $z$, because $z$ is chosen outside of A-B-D past v-cones. Similarly for the A-C-D 3-party marginal. Consequently, it must be either A-B-C that depends on Dave's input $w$ or B-C-D that depends on Alice's input $x$ (or both). Both cases are similar; let us thus consider the case that $p(b,c,d|y,z,w, x)$ depends explicitly on $x$. This is signalling from Alice to Bob-Charlie-Dave. Moreover, this can be used for faster than light communication: it suffices that Bob and Charlie send (at the speed of light) their inputs $y,z$ and outcomes $b,c$ to Dave so that Dave can evaluate their 3-party marginal B-C-D. Since this marginal depends on Alice's measurement setting choice $x$, Alice can communicate to Dave. Figure 4 shows that this communication can be faster than light. By moving B-C-D away from A, but such that the hidden influence from Alice still arrives on time to all of them, one can make this faster than light communication tend to the speed $v$ of the hidden influence.

In summary, the hidden influence of any $v$-causal explanation of quantum correlation can never remain hidden: it necessarily allows for faster than light communication. We'll come back to this remarkable conclusion in section \ref{nosignalling}.

\begin{figure}
\centering
\includegraphics[width=9cm]{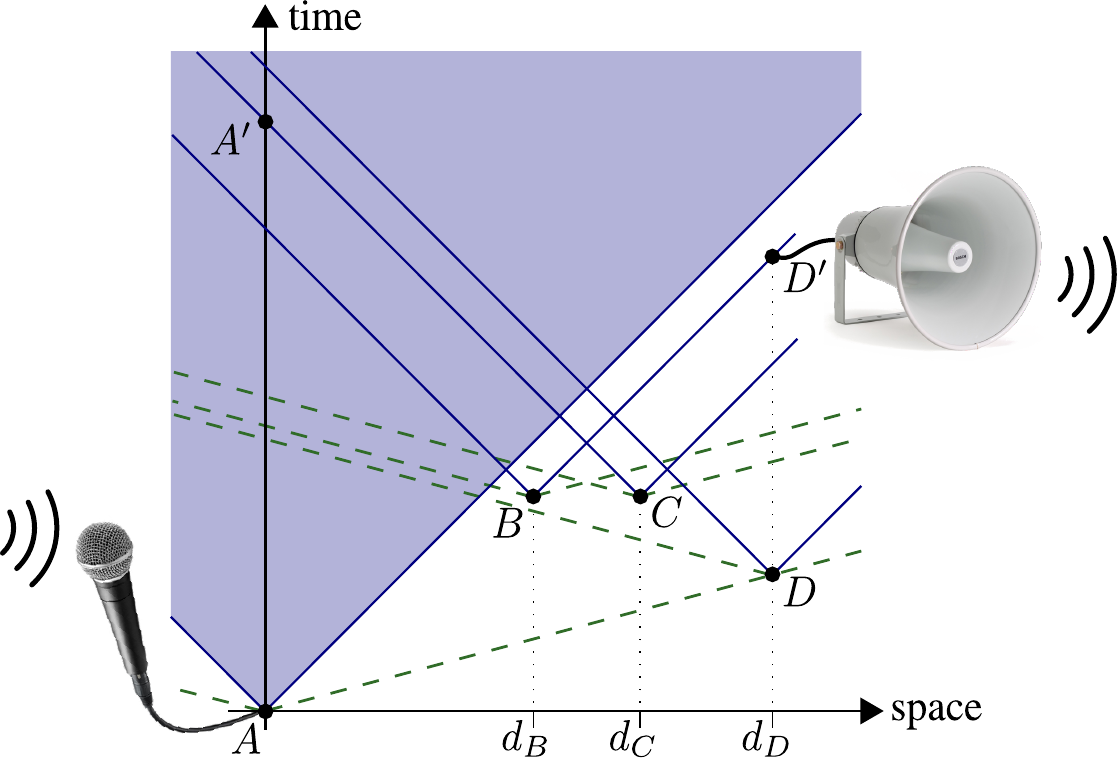}
\caption{In the 4-party scenario, signalling leads to faster than light communication. Here we illustrate the case where the signalling goes from A to BCD; the case D$\rightarrow$ABC is similar (the other cases don't happen, because they are quantum, see text). (Taken with permission from Nature from \cite{Pironioetal}).}\label{fig4}
\end{figure}

\section{Experiment}\label{experiment}
In this section we consider how experiments could test the contradiction we have established between quantum theory, $v$-causality and no faster than light communication. At first, one may wonder whether such an experiment is necessary at all. Indeed, quantum correlations have been measured abundantly. Hence, it seems highly likely that the state (\ref{Psi}) and the quantum measurements (\ref{settings1})-(\ref{settings3}) can be realized with good enough approximation to violate inequality (\ref{ineqS}). Moreover, the very assumption that quantum correlations are explained by $v$-causality implies that the ABD and the ACD correlation predicted by any $v$-causal model are identical to the quantum prediction, hence that any $v$-causal model violates the inequality. If not, the $v$-causal model would not be an explanation for the quantum correlation\footnote{Though, if quantum theory is falsified, then one would no longer be looking for an explanations of all quantum correlations.}. Furthermore, if it would turn out impossible to violate inequality (\ref{ineqS}), then quantum theory would fail even in cases were the events are $v$-connected. This would be very difficult to explain and $v$-causality might not be of much help. This is in sharp contrast to Bell's inequality: had it turn out impossible to violate Bell's inequality, local CC would have been vindicated. Hence, whether or not one eventually observes a violation of the inequality (\ref{ineqS}), in both cases explanations based on  $v$-causality seem difficult to maintain!

But, physics being an experimental science, one should check that correlations violating the inequality we used in the previous section to derive our conclusion can indeed be realized.

So, imagine a source producing a state close to the 4 qubit state (\ref{Psi}) and distributing each qubit to Alice, Bob, Charlie and Dave. Alice is first to choose her measurement setting $x$, measure her qubit and collect her outcome $a$. Alice and her qubit may be aware of the locations of her partners, who may perform some measurement, or are measuring quasi simultaneously, such that the corresponding hidden influence did not reach her yet. However, Alice can't know when such possible measurements will be performed by her partners, or whether they will be performed at all (the so-called ``free-will" assumption). Accordingly, Alice (more precisely her qubit) has to send out her hidden influence at speed $v$ independently of when Bob, Charlie and Dave may measure (or not measure) their qubits.

Dave should be second to measure, but not too early, so as to make sure that Alice's hidden influence reaches him on time. This can be guaranteed by merely letting Dave measure his qubit at a time such that even light would arrive on time. Since the hidden influence propagates faster than light, it will necessarily also arrive on time, irrespectively of which reference frame is the privileged one. Here we assume that Alice's hidden influence always propagates at the same speed $v$, independently of the protocol of the experiment. This is a standard assumption in science: one never assumes that the experimental protocol changes the laws one is testing. In summary, it is easy to guarantee that Dave measures far enough in the future to respect the time ordering of figure 3.

 This just leaves Bob and Charlie. They should both measure in the future of Dave so that its hidden influence arrives on time. This can again be achieved by setting Bob and Charlie in the future light cone of Dave (and thus also of Alice). However, according to the configuration depicted in figure 3, Bob and Charlie should be well enough synchronized to guarantee that no hidden influence from one can reach the other. This is impossible without knowing an upper bound on the speed of the hidden influence and the privileged frame. This difficulty is circumvented, as already explained in the prevous section, by the observation that in the inequality (\ref{Sineq}), no term involves both Bob and Charlie. Hence, one doesn't need to ever measure them in a same round of the experiment. It suffices that, after Dave measured his qubit, a random choice is made by the experimentalist to measure either Bob or Charlie's qubit. In each case, another, fourth (independent) random choice is made to select the measurement setting. Both these choices are made in the future light cone of Dave. Again, we assume that the qubit chosen to be measured, whether it is Bob's or Charlie's, produces a result that is independent of the protocol. In other words, in case Bob's qubit is chosen to be measured, the probability of the result $b$ is the same as if Charlie's qubit would be measured simultaneously: Bob's result probability can't depend on when Charlie's qubit is measured as long as Charlie qubit can't influence Bob's. And if Charlie's qubit is not measured at all, Bob's result probability can't depend on ``when Charlie's qubit is not measured".

In summary, a first random bit decides Alice's measurement setting $x$, next in the absolute future a second random bit chooses Dave's setting $w$, finally, again in the absolute future, a third random bit decides whether Bob's or Charlie's qubit is measured and a fourth random bit decides the measurement setting $y$ or $z$. Note that all these 4 random bits must be independent of the hypothetical additional variables and hidden influences, as in Bell inequality analysis (this is sometimes called the ``free will" or the ``measurement independence" assumption \cite{Hall10,BarrettGisin11}).

In this way, the experimental test of quantum predictions for the configuration depicted in figure 3 can be realized. If a violation is observed, as one expects from quantum theory, then one has to conclude that
\begin{enumerate}
\item either the hypothetical hidden influence can't remain hidden, but necessarily leads to signalling and to faster than light communication,
\item or, all $v$-causal explanations are ruled out, i.e. no combination of Direct Cause and local Common Cause can explain the experimental result.
\end{enumerate}
Both these alternatives are fascinating and will be discussed in the conclusion section.

One might be surprised that the proposed experiment doesn't involve any space-like separated measurements. But, as mentioned at the end of section \ref{vcausality}, the speed of light doesn't appear in the definition of $v$-causality. Hence, according to $v$-causality, one doesn't expect any difference when measurements are time-like or space-like separated. Furthermore, signalling between time-like separated events would be about as bizarre as between space-like separated events. Indeed, imagine that Alice is located in a safe, e.g. in the basement of the Swiss national bank. One expects that this would not affect the correlations between her measurements and those of her partners, wherever they are located. In particular they could be in the future light-cone of Alice, somewhere outside of the bank. But then, signalling from Alice to BCD, as $v$-causality and the violation of (\ref{ineqS}) predict, implies that Alice could communicate to her partners, whatever physical security measures and isolation one imposes on Alice\footnote{This would be similar to signalling using gravitation - no way to prevent it - but at the speed $v$.}!

\section{Newton and the principle of continuity}\label{Newton}
It is not the first time in history that physics is confronted with nonlocality. Actually, physics almost always presented a nonlocal world-view of nature, first with Newton's theory of universal gravitation, then with quantum nonlocality. Only during short time window of about 10 years did physics present a local world-view.

Newton was very concerned by the nonlocal predictions of his theory of universal gravitation. Indeed, he noticed that his theory predicts that any change in the local configuration of matter would have an immediate effect on the entire universe. Hence, by moving to the left or to the right a stone on the moon\footnote{To move the stone one shouldn't take support on the moon, as this would not move the center of mass of the moon-\&-stone, but use a small rocket.}, one could, in principle, signal at arbitrary speed to Earth and to any place in the universe. Let us read how the great man described the situation: \cite{NewtonNonlocality}:

{\it That Gravity should be innate, inherent and essential to Matter, so that one Body may act upon
another at a Distance thro' a Vacuum, without the mediation of any thing else, by and through which
their Action and Force may be conveyed from one to another, is to me so great an Absurdity, that I
believe no Man who has in philosophical Matters a competent Faculty of thinking, can ever fall into
it. Gravity must be caused by an Agent acting constantly according to certain Laws, but whether
this Agent be material or immaterial, I have left to the Consideration of my Readers. }

Accordingly, ``no action at a distance" is not a principle of relativity nor of Einstein, but is part of Newtonian space-time. Let us emphasize that ``no action at a distance" implies that nothing propagates at infinite speed, in particular there are no infinite speed influences nor $\infty$-causality.

Usually, quantum correlations are seen as being in tension with (special) relativity, remember Shimony's statement about the peaceful coexistence of quantum theory and relativity. But it is natural to go beyond these tensions and investigate the consequences of assuming that the correct interpretation of Lorentz transformation is not mere geometry of space-time, but real Fitzgerald contractions of lengths and Larmor dilation of time intervals, as Lorentz and Poincar\'e themselves thought and as John Bell considered \cite{BellTeachRelativity,BellHiddenComm}. Hence, the interest for studying quantum correlations in Newtonian space-time, or, equivalently for that matter, in a universal privileged reference frame.

Notice that all $v$-causal explanations of correlations satisfy a {\bf principle of continuity} that states that everything (mass, energy and information) propagates gradually and continuously through space as time passes, i.e. nothing jumps instantaneously from here to there. In other worlds, there is no action at a distance. Reciprocally, all explanations of correlations that satisfy the principle of continuity are $v$-causal. Hence, Newton and Einstein would have bet on a $v$-causal explanation of all correlations, including quantum correlations.

An experimental violation of the inequality $S\le 7$ either implies a violation of the principle of continuity or implies faster than light communication.

\section{No-signalling in $v$-causality}\label{nosignalling}
No-signalling is generally considered as a fundamental principle that has to hold in any meaningful physical theory.
However, if the correlations between some events are due to hidden influences, then there is no reason to assume that the influences don't allow one to signal (at the speed of the hidden influence or slower). This is for example the case with gravity. Had someone before Einstein had the technology to check the correlation between the displacement of a stone on the moon and the weight of some mass on Earth, even when displaced and measured simultaneously, he would have observed a null correlation (at least for good enough synchronization) and thus have falsified Newton's theory of universal gravitation. He could also have observed that the correlation establishes when the weight measurement is performed about a second after the displacement of the mass on the moon. This would have allowed him to signal at the speed of what was then a hidden influence, i.e. the speed of gravitons that, according to general relativity, carry the cause of the change in the weight of the mass on Earth. This is a typical Direct Cause explanation. Note that one could have used this hidden influence to signal even without knowing the theory of general relativity.

Similarly, if the speed of the hidden influence that explains quantum correlations propagates faster than the speed of light, then the corresponding signalling would equally be faster than the speed of light. Consequently, there are only two possibilities:
\begin{enumerate}
\item either the hidden influence remains hidden for ever, i.e. is intrinsically hidden\footnote{I am quite suspicious of explanations relying on intrinsically hidden stuff, hence I dislike this part of the alternative.}, hence doesn't allow for signalling, or
\item the hidden influence can be used to communicate at a speed equal or lower than the speed of the hidden influence, i.e. the hidden influence doesn't remain hidden.
\end{enumerate}
In section \ref{NoInfluenceNorLocalVariables} we demonstrated that the hypothetical hidden influence of all $v$-causal model can't remain hidden, but on the contrary leads to faster than light communication at the level of the classical measurement settings and results. Hence, the first of the above two alternative is excluded.

\section{Conclusion}\label{conclusion}
The main conclusion of this paper is that an experimental violation of the inequality $S\le 7$ would imply
\begin{enumerate}
\item either a violation of the principle of continuity (that states that everything propagates gradually and continuously through space as time passes as discussed, in section \ref{Newton}), i.e. the falsification of all $v$-causal models, or
\item the possibility of faster than light communication at the level of the classical measurement settings and results.
\end{enumerate}

It is unlikely that many physicists will contemplate seriously the second alternative\footnote{One recent exception is B. Cocciaro \cite{Cocciaro12}. In this paper the author also recalls that faster than light communication in one universal global privileged reference frame, as consider in this paper, doesn't lead to the ``grand father" time paradox. Indeed, for time paradoxes one should communicate to one's own past; this requires a go-\&-return communication. But if both the go and the return signal are defined in the same reference frame and at the same - possibly supraluminal - speed $v$, then the ``return" signal will necessarily arrive in the absolute future of the start of the ``go" signal. It is straightforward to see this in the privileged frame. But then, the start of the ``go" and the arrival of the ``return" signals are necessarily also time-like in all other reference frames, hence the impossibility to communicate to one's own past. This is not new and was emphasized, e.g., in  \cite{maudlin,Reusse84,Caban99,Cocciaro12}. Consequently, supraluminal communication might not have said it's last word.}. However, one should realize that the first alternative is about as difficult to swallow as the second one. A violation of the principle of continuity implies that the world is truly and definitively not local, i.e. Nature is not continuous, but nonlocal. This conclusion has already been claimed by many physicists (including this author), though only based on the violation of Bell inequality between space-like separated events. These physicists made the (admittedly highly plausible) assumption that space-time is described by relativity. In this paper we have extended the conclusion: even if one is willing to consider a Newtonian-type privileged reference frame, but without faster than light communication, the conclusion that Nature is nonlocal is unavoidable.

Should then Physicists give up the great Enterprize of explaining how Nature does it \cite{GisinScience}? Certainly not! But physicists have only two options:
\begin{enumerate}
\item Pursue the search for the speed of $v$-causal explanations by improving the "Salart-type" experiments \cite{Salart,Cocciaro10}. Note that the finding of such a speed would falsify both quantum theory and relativity, a result not many physicists are willing to envisage. However, the tension between these two pillars of today's physics may well dissolve not merely by saving one of them at the cost of the other, but by finding the limits of both theories. Accordingly, "Salart-type" experiments are still needed, but the result of \cite{Pironioetal} -- recalled in this paper -- shows that a positive result would definitively lead to faster than light communication, hence it would not only falsify quantum theory but also falsify relativity.
\item Accept quantum nonlocality and enlarge our story tool-box by inventing new tools -- necessarily nonlocal -- to tell explanatory stories. Possibly something like ``one random event can manifest itself at several locations".
\end{enumerate}

\small
\section*{Acknowledgment}
This article greatly profited from numerous exchanges with my co-authors of \cite{Pironioetal} and from comment by Rob Thew and many colleagues over the years.
This work has been supported by the ERC-AG grant QORE,  the CHIST-ERA DIQIP project, and by the Swiss NCCR {\it Quantum Science and Technology} - QSIT.


\end{document}